# MIGRATION OF DATA FOR IKNOW APPLICATION AT EURM - A CASE STUDY

Marko Vučković     Toni Stojanovski
Faculty of Informatics, European University
Skopje, Republic of Macedonia

ABSTRACT

**Software evolves. After many revisions and improvements software gets retired and replaced. When replacement takes place, one needs to migrate the data from the old database into the new database, so the new application can replace the old application. Student administration application (SAA) currently used by European University (EURM) has been outgrown by the university, and needs replacement. iKnow application developed as part of the iKnow Tempus project is scheduled to replace the existing Student Administration application at EURM. This paper describes the problems that were encountered while migrating the data from the old databases of SAA to the new database designed for the iKnow application. The problems were resolved using the well-known solutions typical for an ETL process, since data migration can be considered as a type of ETL process. In this paper we describe the solutions for the problems that we encountered while migrating the data**.

## I. INTRODUCTION

Many institutions today use some sort of application for employing the business rules and managing the data. The usage of this type of applications can provide many advantages for the institutions. It can reduce the human error, speed up the information retrieval, strengthen the business rules, and at the same time provide storage and management of large amounts of data. This large amount of data is stored in a database specifically designed for the application and the type of institution that will use the application. At some point of time, the institution may require to replace the using application with new application. There can be many reasons for this requirement. Some of them are: the new system can be more reliable, it can provide new functionalities, improve the data integrity, work faster, may cut the hardware and network requirements etc.

European University - (EURM) is a private university based in the Republic of Macedonia. The application that is currently used at EURM was designed to fit the requirements of educational institution and therefore provide system for managing employees, students, programmes, payments, classes, exams and any other entities that describe one educational institution. EURM is a partner institution in the iKnow Tempus project [1]. EURM plans to replace its existing custom-made Students Administration software application (SAA) with the iKnow software which will be delivered by the iKnow project. It is expected that the iKnow software will further automate the business processes at EURM, will provide functionalities and access to new groups of users e.g. professors, while providing improved performance at lower total cost of ownership.

When this type of application change is needed, the data that was stored in the old application database needs to be migrated to the database that is used by the new application. This migration will allow the old application to be replaced with the new application. SAA application and iKnow application are both using MSSQL server databases. SAA uses three different databases for storing the data needed for the three different study cycles, while iKnow application stores the data for the three study cycles in one database. The problems that will be discussed in this paper are caused by the different designs that these two databases have.

There are problems that are inevitable when migrating data between databases with different designs. Some of the problems that occur are described in [3], like missing data (appearance of blank fields, NULL fields or defaults), mismatched data (field overuse) or poor quality data (appearance of deficiencies). The purpose of this paper is to point out the problems that were encountered while migrating the data from the SAA database to the iKnow database, and to explain the methods we have used to resolve these problems.

Clearly, data migration process is only one challenge of many that exist when migrating to a new application. Once the data migration is done, a pilot group of users e.g. academic staff and students from one faculty only, will start using the application. Only after a successful pilot phase, the application can be put into full production. Pilot phase, training of users, and change management are out of scope for this paper.

The outline of the paper is as follows. In section 2 we give brief explanation about the iKnow application and the SAA application currently used at EURM, and we introduce the process that was used for the migration. In section 3 we explain the solutions of the problems that were encountered while migrating the data. In section 4 we give our contribution by improving the scalability of the iKnow application by modifying some of the stored procedures that this application uses. Section 5 concludes this paper.

## II. BACKGROUND

During the past 10 years EURM has been using a custom-made application to automate a set of business processes. There are several issues related to this application: (i) many business processes were left out, and as a consequence paper based documents are being extensively used in the everyday operations at EURM; (ii) data entry is almost exclusively done by the student administration (SA) staff, which puts too much burden on the SA staff; (iii) Academic staff and students have almost no access to student records, exam records etc. (iv) Separate databases exist for each cycle of studies, which duplicates the data entry process.

iKnow application is expected to resolve the above mentioned problems: More business processes will be covered by iKnow. Academic staff and students will be active participants in many business processes, which will reduce the amount of work done by SA staff. SA staff in turn can provide higher quality service to students. Multiple data entry will be eliminated thus reducing data errors.

As mentioned before, current EURM application SAA uses three databases. This is because the application is using different database for every study cycle available on the university. The first database is for the graduate study cycle, the second is for the one year postgraduate study cycle, and the third one is for the two year postgraduate study cycle. We are faced with the task of migrating the data from the old databases to the new database. Data migration from the old databases to the new database from the iKnow software is a typical example of an *Extract, Transform, and Load* (ETL) process. Next subsection gives an overview of ETL, and the typical problems of an ETL process. Then it is followed by a detailed description of the specific ETL problems we encountered during the migration, and our approach to solving them.

### A. Extract, transform, and load (ETL) process

Extract, transform, and load (ETL) is a process in data migration e.g. from legacy to new applications that includes ([2], page 125):
- Extracting data from legacy database
- Transforming it to fit the data schema of the new database
- Loading it into the new target database

*Extract* is the first part of an ETL process which involves extracting the data from the legacy databases, and converting it into a single format which is appropriate for transformation processing. Each separate legacy database may use a different data format, which calls for data consolidation during the extract stage. Extraction parses the extracted data, and checks if the data meets an expected pattern or structure. Non-compliant data may be rejected in full or partially.

The *transform* stage applies a series of rules or functions to the extracted data from the source to derive the data for loading into the end target. Following transformation types are frequently required to meet the functional and technical needs of the target database:

P1. Selecting only certain columns to load
P2. Translating coded values (e.g., 1 for male and 2 for female from the legacy system is translated into M for male and F for female in the new database)
P3. Deriving a new calculated value (e.g., *afterTax* = *net* * (1+*tax*))
P4. Filtering
P5. Sorting
P6. Joining data from multiple sources (e.g., lookup, merge)
P7. Generating surrogate-key values
P8. Splitting a column into multiple columns
P9. Splitting a table into two or more related tables
P10. Data validation

Should an exception occur during any transformation, it should be dealt with appropriately e.g. data is disregarded, or ETL process is aborted.

The *load* phase loads the data into the target database. The constraints defined in the target database schema are applied in this stage e.g. uniqueness, referential integrity, allowed values. This ensures the overall quality of the migrated data, and their compliance with the business rules enforced in the database schema.

### III. OUR ETL PROCESS

The key requirement is to fully understand the two database designs. This will help us to identify the related tables and fields from the old database and the new database, and to migrate all records from the old database into the matching table in new database while preserving all relations between the records from different tables. This makes the transformation stage the most important stage of the ETL process. The methods used for transformation greatly depend on the database schema and relations.

Here is the list of problem we encountered during the ETL process:
- *Problem 1*: The foreign key constraint, which enforces referential integrity [4], does not allow us to insert a value for a foreign key field if the foreign key table does not have a record with the corresponding primary key. One needs to define a table order by which the tables should be populated.
- *Problem 2*: Several tables that exist in the iKnow database: *Nationality*, *Community*, *Countries* etc. do not have corresponding tables in the SAA databases. The information that describes the nationality, birth or living community, and countries in the SAA databases can be found as fields in tables, mostly as *nvarchar* data type.
- *Problem 3*: There are situations when there is a need for keeping the existing IDs for specific table records. This can simplify the transformation stage methods. Since there are three SAA databases with exactly the same schema, existing IDs could have been be kept only if the three tables from the three SAA databases contained the same records, which is not the case.
- *Problem 4*: How to distinguish records that are coming from the same table, but from different SAA database? How to divide those records into different study cycles?
- *Problem 5*: Certain tables such as *Students*, *Programmes*, *Courses* etc. exist in the three databases but contain different records. Records with same *ID* exist in the same table in different SAA databases, but contain data for different entity e.g. student from different study cycles. We use mapping tables to overcome this problem.

The solutions that we used for these problems are described in the following subsections.

### A. Problem 1

Priority of tables during the migration process was determined starting from this rule: Tables that do not contain foreign key fields has the highest priority. The tables with the highest priorities are the first to be migrated and populated. If a table has one or more foreign key fields, then the priority of that table is lower than the priority of its foreign key tables. The "populate by priority" method will preserve the foreign key constraint.

### B. Problem 2

We use the following approach to solve this problem. Distinct data contained in free-text fields from the three SAA databases are extracted and then, typically without any transformation, and then are loaded into the corresponding table in the target database. Assume that the nationality information is stored for every student, teacher and applicant as *nvarchar* field, into the SAA databases. We extract these values, as distinct values (*SELECT DISTINCT*), from the three SAA databases, and populate the *Nationality* table located in iKnow database. Thereby every distinct nationality gets its own ID number, which will be used as foreign key in other tables. Same solution is applied for the similar tables such as *Community*, *Countries* etc.

### C. Problem 3

*Faculty* table exists in every SAA database. The faculties are the same for every study cycle and therefore for every SAA database. But over time some rows from these tables were deleted. Because of this the ID field values, which were automatically incremented integer values, are now inconsistent. If we try to load the extracted faculties into the corresponding table from the iKnow database, then new auto incremented integer values will be inserted into the faculty ID field. This must not happen. Our solution is to temporarily turn off the auto increment operation on the ID field for the target database table and insert the existing IDs from the old database table. Code 1 explains this solution in more details.

```
#allows inserting values in auto incremented field
SET IDENTITY_INSERT iKnow.Faculty ON

#Load the extracted records
INSERT INTO iKnow.Faculty
SELECT * FROM SAA.Faculty

#restore auto increment operation
SET IDENTITY_INSERT iKnow.Faculty OFF
```

Code 1. Pseudo code for inserting with IDENTITY_INSERT

*SET IDENTITY_INSERT <table name> ON* command allows values to be inserted into fields that have auto increment property. After the insertion, *SET IDENTITY_INSERT <table name> OFF* command instructs the database to automatically increment the *ID* field for newly added records. *ID* field is incremented starting from the field's highest existing value.

Because the source *Faculty* table stores the same information for every SAA database, only records from one SAA database *Faculty* table needs to be extracted and loaded.

### D. Problem 4

iKnow database has table called *StudyCycles* that is used to distinguish the different study cycles, unlike the SAA application which uses different databases per study cycle. At the start of the migration, we inserted records (study cycles) into *StudyCycle* table that correspond to the SAA database. The table records from the iKnow database tables that differ per study cycle have foreign key to the *StudyCycle* table.

### E. Problem 5

A mapping table is a temporary table used to store the mapping between ID identity fields in related tables from the old and new database. Mapping tables are used only during the migration process. Figure 1 explains the mapping tables.

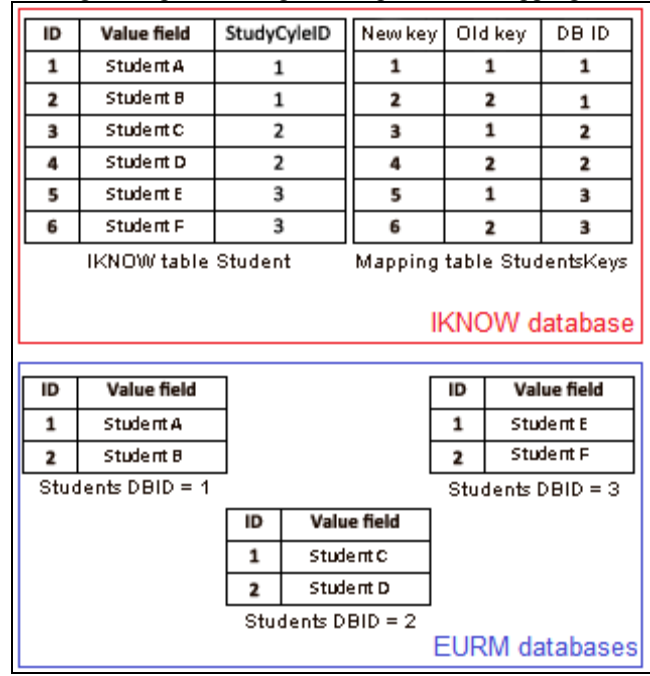

Figure 1. Usage of mapping table (example for *Student* table).

A mapping table consists of three integer fields: *NewKey* (stores the new IDs generated when populating the target iKnow table), *OldKey* (stores the old IDs from the SAA databases) and the *DBID* field (stores a value that distinguishes between the three SAA databases). Once the ETL process is finished, a target table from the iKnow database will store all records that were previously divided amongst the identical tables in the three SAA databases. The mapping table is used by tables containing foreign key relationship to translate the old values for the foreign keys into new values for foreign keys. Even though Figure 1 shows an example mapping table for the *Student* table, the same table schema is used for the mapping tables *ProgrammesKeys*, *CoursesKeys* etc. for tables like *Programmes*, *Courses* etc. We illustrate the mapping tables through the *ProgrammesCourses* table.

The iKnow *ProgrammesCourses* table represents the relation between the iKnow *Programme* table and the iKnow *Course* table. The "populate by priority" rule suggests that the *ProgrammesCourses* table should be populated after *Programme* table and the *Course* table. During the migration of *Programme* and *Course* tables, two mapping tables are created: *ProgrammesKeys* and *CoursesKeys*. Migration of *ProgrammesCourses* table process starts with extraction of the needed data and applying transformation methods such as joining (P6). Join is needed between the source *ProgrammesCourses* table from the SAA database, and the newly created mapping tables. The join is done on the source table's *ID* field and the mapping table's *OldKeys* field, and also uses the *DBID* field. This join is needed so that we can identify the new values for the ID fields in *Programme* and *Course* tables, and insert the new keys into the *ProgrammesCourses* table from the iKnow database. The process is repeated for every SAA database.

## IV. INCREASING THE SCALABILITY AND PERFORMANCE

The iKnow application is an ASP.NET application that uses ASP.NET Membership Provider for managing users and their roles. In previous studies [5][6] we identified several improvements that can be applied to the membership provider.

These improvements are based on improving the response time of the stored procedures that are used by the membership provider. SQL stored procedures from the membership provider use paging technique for dividing the large result sets into smaller result sets that can be displayed in a web page. Our research [6] analyses different SQL server-based paging techniques, and identifies a design for the SQL stored procedures that can improve the response time, scalability and performance of the ASP.NET Membership Provider.

Figure 2 gives the execution time vs. page index for 100.000 user accounts. Execution times are calculated as an average value over 30 executions of a stored procedure for each tested value of the page index. Page size is fixed to 10.

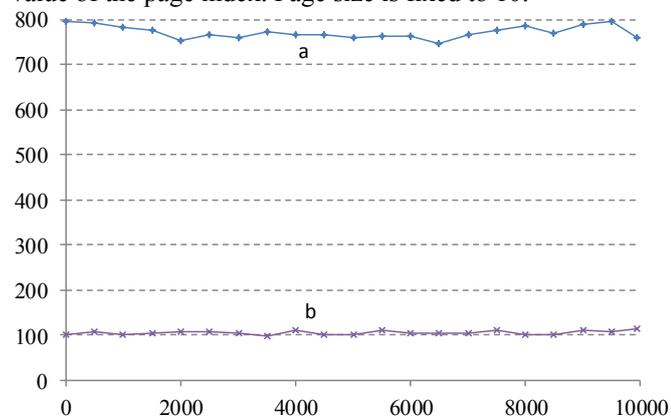

Figure 2. Execution time [ms] of stored procedure aspnet_Membership_GetAllUsers vs. page index for 100.000 records. a) Original. b) Improved and sorting by LoweredUserName.

Curves a) and b) in Figure 3 depict the dependence of the average execution time for the original and our improved version of aspnet_Membership_GetAllUsers, respectively, on the number of user accounts.

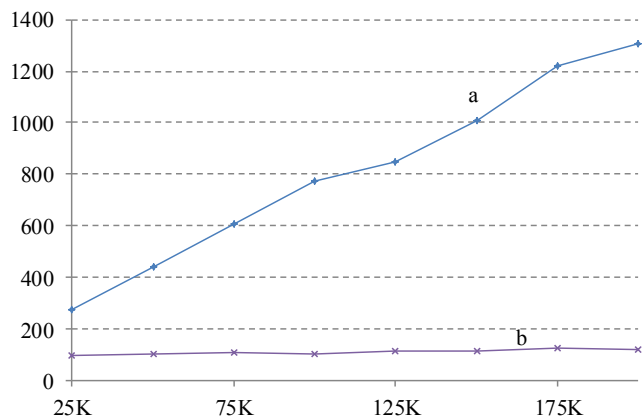

Figure 3. Execution time [ms] of stored procedure aspnet_Membership_GetAllUsers vs. number of records. a) Original. b) Improved and sorting by LoweredUserName.

The iKnow ASP.NET application uses the default SQL stored procedures from the ASP.NET Membership Provider. By using our improved procedures for paging the results, the iKnow application can increase the scalability and its performance.

## V. CONCLUSION

The data migration from the old EURM application's databases to the new iKnow application database is essential for the iKnow application to go through a pilot phase, and then to become fully operational. Data migration is a type of ETL process, and in the migration of the old data to the iKnow database we applied solutions typical for the ETL process.

The main purpose of the data migration is ensuring the overall quality of the migrated data.

By implementing the improvements from our previous studies, the iKnow application can increase the scalability and its overall performance.

The methods that were used for this data migration can provide guidelines for other data migrations and can help other educational institutions that are partners of the Tempus project to migrate their data to the iKnow application.

## REFERENCES

[1] Project Tempus JPGR 511342 – iKnow, http://iKnow.ii.edu.mk/
[2] William H. Inmon, Bonnie O'Neil, Lowell Fryman, "Business Metadata: Capturing Enterprise Knowledge"; Morgan Kaufmann, 1st edition 2007, ISBN-10: 0123737265.
[3] Data migration management, A methodology: Sustaining data integrity after the go live and beyond, UTOPIA Inc., 2009
[4] FOREIGN KEY Constraints, MSDN (Visited on 15 March 2012) URL: http://msdn.microsoft.com/en-us/library/ms175464.aspx
[5] T. Stojanovski, M. Vučković, I. Velinov, "Empirical study of performance of data binding in ASP.NET web applications", ETAI Conference, September 2011, Ohrid, Macedonia (http://arxiv.org/abs/1201.0357v1).
[6] Toni Stojanovski, Ivan Velinov, Marko Vučković, "Scalability of Data Binding in ASP.NET Web Applications", submitted to WSEAS TRANSACTIONS on Computers, http://arxiv.org/abs/1202.3255